\documentclass[aps,twocolumn,prb,superscriptaddress,noshowpacs,nofootinbib,noshowkeys,floatfix]{revtex4-2}
\usepackage[colorlinks=true,linktocpage=true,linkcolor=blue,citecolor=blue]{hyperref}
\usepackage[usenames,dvipsnames]{color}
\usepackage{amsmath, amssymb}
\usepackage{multirow}
\usepackage{longtable}
\usepackage{comment}
\usepackage{color}
\usepackage{xcolor}
\usepackage{xspace}
\usepackage{cleveref}
\usepackage[normalem]{ulem}  
\usepackage{graphicx,lipsum}
\usepackage{ragged2e}
\setcitestyle{numbers}
\setcitestyle{square, ulem}
\begin{document}

\title{Half-metallicity and wandering axis ferromagnetism in Fe$_2$Ti$_{1-x}$Mn$_x$Sn (0 $< x \leq$ 0.3) Heusler Alloys}
\author{Kulbhushan Mishra}
\affiliation{Indian Institute of Technology, Khandwa Road, Indore, Simrol, 453552, India}
\author{Shishir Kumar Pandey}
\affiliation{Birla Institute of Technology \& Science, Pilani Dubai Campus, Dubai International Academic City, P.O. Box No. - 345055, Dubai, UAE}
\author{S. Chaudhuri$^{\dag}$}
\affiliation{Center for Condensed Matter Science, National Taiwan University, Taipei 10617 Taiwan}
\altaffiliation[Current Address of SC:~]{GITAM University, Rudraram, Hyderabad, Telangana, 502329, India.}
\author{Rajeev Rawat}
\affiliation{UGC-DAE Consortium for Scientific Research, University Campus, Khandwa Road, Indore 452001, India}
\author{P. A. Bhobe}
\email{pbhobe@iiti.ac.in}
\affiliation{Indian Institute of Technology, Khandwa Road, Indore, Simrol, 453552, India}

\begin{abstract}
We investigate the effect of Mn substitution in Fe$_2$Ti$_{1-x}$Mn$_x$Sn on electronic structure and magnetic and electrical transport properties. The spin-polarized density of states calculations using density-functional theory (DFT) yields a half-metallic ground state in Mn-rich compositions. Localized magnetic moments at Mn sites interacting through the cloud of conduction electrons formed by Fe and Ti atoms are also predicted. Electrical resistivity and magneto-transport measurements reveal a Kondo-like ground state at low temperatures and a peculiar linear negative temperature coefficient of resistance in the high-temperature regime with a predominant electron-phonon scattering mechanism. Analysis of room temperature powder X-ray diffraction data reveals a highly ordered L2$_1$ structure and reduction of antisite disorder upon Mn substitution. The temperature-dependent magnetization measurements reveal distinct features indicative of weak anisotropy in the system. Isothermal magnetization measured as a function of the applied field helps identify the unique magnetic ground state of the half-metallic Fe$_2$Ti$_{1-x}$Mn$_x$Sn composition as a ferromagnet with a wandering axis that distinctively orients in the direction of the applied magnetic field. The measurement of X-ray absorption fine structure (XAFS) reveals that the random anisotropy arises due to the local lattice distortion around Mn atoms in the prepared compositions. Our findings thus provide a new perspective for studying the mechanism of half-metallicity and associated magnetic order in Heuslers.
\end{abstract}

\maketitle
\section{Introduction}
Since its discovery in 1903\cite{FHeusler}, Heusler alloys with general formula X$_2$YZ (where X and Y are transition metals and Z is the main group element) have constantly drawn the attention of the researchers due to their remarkable properties, such as magnetic shape memory effect\cite{SME}, near - room temperature magnetocaloric effect\cite{PAB_apl}, half-metallic ferromagnetism\cite{HM1, HM2}, spin-gapless semi-conductivity, unconventional superconductivity\cite{SC_heusler} and anomalous Hall effect, and so on. These compositions crystallize in the L2$_1$ structure (space group: \textit{Fm-3m}) consisting of 4 interpenetrating FCC sublattices, where X atom occupies the (0.25,0.25,0.25) and (0.75,0.75,0.75) sublattice sites, whereas the Y and Z atoms occupy the (0.5,0.5,0.5) and (0,0,0) sublattice sites, respectively. Theoretical studies on Heusler alloys reveal an interesting relationship between the total spin moment per unit cell (M$_T$) and the number of valence electrons count (VEC) per formula unit (Z). Also known as Slater-Pauling rule\cite{SPrule}, this relationship states quantitative value of M$_T$ as equal to (Z - 24), and seems to provide a handle to modulate the physical properties of Heuslers through suitable chemical substitutions. 

Ideally, for compositions where VEC equals 24, the Slater-Pauling rule suggests a non-magnetic ground state. Examples include the Fe$_2$VAl, Fe$_2$VGa\cite{Fe2VAl, Fe2VGa} compositions. However, measurement of the magnetic properties of these compositions show signatures of weak magnetism\cite{Fe2VAl2, Fe2VAl3}, the reason being the presence of anti-site and quenched disorders\cite{Fe2VAl_quenched_disorder}. Heusler alloys with VEC 24 are also considered for applications such as magnetic information storage, where they could play the role of a thin non-magnetic buffer layer sandwiched between two ferromagnetic layers of other Heusler compositions, acting as a spin polarizer in a recording head while providing the very compatible inter-layer coupling. Furthermore, a significant decrease in the resistivity, magnetization, and magnetoresistance values has been reported when other transition metal atoms are substituted at X or Y sites, which can lead to a state of high spin-polarization\cite{Fe2VAl_doping} at Fermi level. A narrow gap/pseudogap forms near the Fermi level, with transport measurements reflecting a semiconducting nature and heavy effective mass of the charge carriers, making them suitable for thermoelectric applications. 

Another notable example is the Fe$_2$TiSn composition, which conforms to the VEC = 24 prototype. The electronic structure calculation indicates Fe$_2$TiSn to be a non-magnetic semimetal with a pseudogap in the density of state at the Fermi level. Based on specific heat measurements, it has been speculated to be a heavy fermion metal with quasi-particle mass $\sim$ 40 m$_e$\cite{FTS_HF}. The substitution of Ni at the Fe site forms a system with isolated spins, leading to a Kondo-like interaction between localized spins and conduction electrons \cite{FTS_kondo}. On the contrary, Co substitution leads to a typical spin fluctuation scenario. Optical conductivity and infrared studies of Fe$_2$TiSn provide evidence of pseudogap in the density of states and the mass enhancement at low temperatures to be Schottky anomaly from magnetic clusters resulting from antisite disorder between Fe and Ti sites. It is to be noted here that in contrast to Fe and Co, bulk Ni has a sharp peak at E$_f$ in one of the spin sub-bands\cite{dosNi}. This sudden increase of DOS at E$_F$ may lead to enhanced electron-electron interaction (EEI), leading to Kondo-like spin scattering at low temperatures. Furthermore, the non-magnetic semi-metallic band structure of Fe$_2$TiSn demonstrates a low density of states at the Fermi level, rendering the electronic structure and magnetic properties highly sensitive to doping or elemental substitution and with proper substitution at Ti and Sn sites, the system can be driven to half-metallicity\cite{sayanJPCM2, sayanSR}.

In this work, we report the detailed investigation of the structural, magnetic, and transport properties of Mn-substituted Fe$_2$TiSn compositions. X-ray diffraction (XRD) results show the formation of an ordered L2$_1$ structure with highly reduced antisite disorder. Temperature and field-dependent magnetization measurements indicate the presence of anisotropy in the system due to the strong local magnetic moment of Mn atoms. The low-temperature upturn in the electrical resistivity is due to the combined effect of electron-electron correlation and strong Kondo-like spin scattering. The $\rho$(T) above 100 K can be best explained by electron-phonon and electron-magnon scattering, and the contribution of electron-magnon scattering is comparatively less, which indicates the presence of spin polarization of density of state in the prepared composition. We have also performed the electronic structure calculations to better understand the ground state properties. 

\begin{table*}[ht]
  \begin{center}
    \caption{ Atomic percentages, parameters determining goodness of fit and magnetic transition temperatures of Fe$_2$Ti$_{1-x}$Mn$_x$Sn. }
    \label{tab: Table_1}
      \setlength{\tabcolsep}{10pt}
    \begin{tabular}{c c c c c c c c c c c c}
    \hline
     Composition & \multicolumn{4}{c}{Atomic $\%$} & \multicolumn{3}{c}{goodness of fit} & T$_{C}$ & T$_{S}$ \\
            & Fe & Ti & Mn & Sn & R$_{Bragg}$ & R$_{F}$ & $\chi^2$ & (K) & (K) \\
         \hline
      0.05 & 48.05 & 24.51 & 1.23 & 26.21 & 8.12 & 10.1 & 1.94 & $<$ 5 & 250\\
      0.10 & 48.29 & 22.76 & 2.55 & 26.27 & 8.44 & 6.49 & 2.51 & 6 & 268\\
      0.25 & 48.0 & 19.55 & 6.25 & 26.13 & 9.68 & 7.86 & 2.77 & 18 & 320\\
      0.30 & 47.91 & 18.10 & 7.40 & 26.60 & 10.7 & 9.09 & 2.02 & 25 & 325\\
      \hline
    \end{tabular}
  \end{center}
\end{table*}

\section{Experimental and Computational details}
\begin{figure}[ht]
    \centering
     \includegraphics[width=\linewidth]{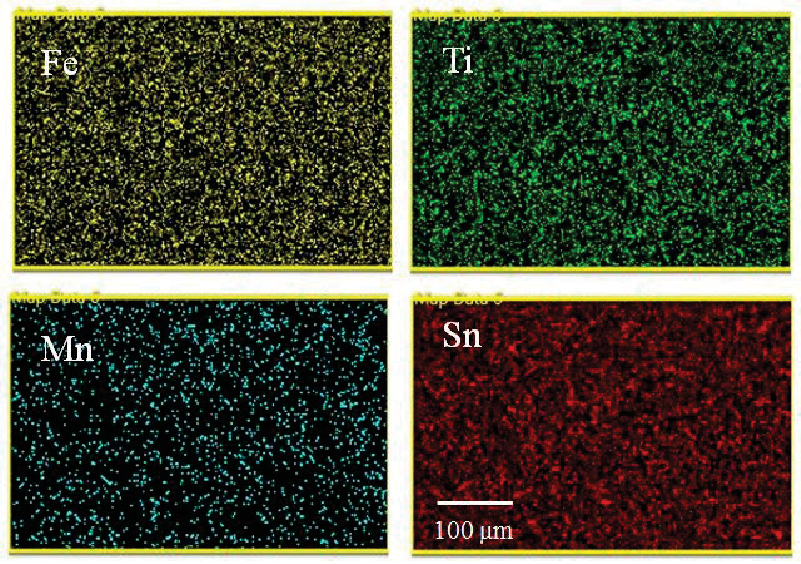}
   \caption{The elemental mapping results of $x$ = 0.25. The figures are presented at a scale of 100 $\mu$m.}
    \label{fig: Fig.1}
\end{figure}

Polycrystalline Fe$_2$Ti$_{1-x}$Mn$_x$Sn compositions were prepared by melting stoichiometric amounts of high purity (99.99\%) elements such as Fe, Mn, Ti, and Sn using an electric arc furnace by placing these elements on a water-cooled Cu hearth and under flowing Ar atmosphere. The homogeneous ingots thus formed were vacuum sealed in the quartz tubes and annealed for 72 hours at 800$^\circ$C followed by quenching in ice water. The weight loss during the whole process was $\leq$ 1$\%$. The energy dispersive X-ray spectroscopy (EDS), equipped with a field emission scanning electron microscope, was performed using Model FE-SEM Supra 55 (Carl Zeiss, Germany) to ensure the target composition after melting was close to the starting stoichiometry. The atomic percentages of the elements are presented in Table-\ref{tab: Table_1}, and the elemental mapping for $x$ = 0.25 is shown in Fig.\ref{fig: Fig.1}. The crystallographic phase purity of the samples was checked using a laboratory X-ray source on Bruker D8 Advance diﬀractometer having Cu target ($\lambda$ = 1.5406 \AA), confirming the formation of the Heusler phase. Magnetization as a function of temperature and magnetic field was measured using a Quantum Design SQUID magnetometer. A standard four-probe method with a homemade resistivity setup equipped with a superconducting magnet system was used for the electrical resistivity and magnetoresistance measurements. The Extended X-ray absorption fine structure (EXAFS) measurement was carried out at P65 beamline, PETRA-III DESY, Hamburg, Germany at Fe (7112 eV) and Ti (4966 eV) K edges at 6 K. Gas-filled ionization chambers were used to simultaneously record the incident ($I_0$) and transmitted ($I$) photon energies. The data was collected over various scans to reduce the statistical noise. Subsequently, the EXAFS signal was extracted and analyzed in the 3 to 14 \AA$^{-1}$ range using Demeter suite\cite{Demeter}.

Density-functional theory (DFT) calculations are performed within the Quantum Espresso package~\cite{qe2, qe1} where ultrasoft pseudopotentials~\cite{usp} and a plane-wave basis set are employed. Generalized-gradient approximation (GGA) with Perdew–Burke–Ernzerhof (PBE) functional form is used for the calculation of exchange-correlation energy~\cite{PBE}. For Sn, Ti, Mn and Fe, 4$d^{10}$5$s^2$5$p^1$, 3$s^2$3$p^6$4$s^2$3$d^1$, $s^2$3$p^6$4$s^{1.5}$3$d^5$4$p^0$ and 3$s^2$4$s^2$3$p^6$3$d^6$ are treated as valence states respectively. The kinetic energy cutoff for plane-wave functions of 95 Ry and the charge density of 1140 Ry are considered in our calculations. We start with the simple cubic crystal structure (space group: $Fm-3m$, no: 225) of Fe$_2$TiSn with lattice constant $a$ = 6.059 \AA{} and fully optimize it after replacing one of the Ti atoms with Mn (25 \% doping). Energy convergence criterion of 5 $\times$ 10$^{-7}$ Ry and 6$\times$6$\times$6 uniform $k$-mesh and Marzari-Vanderbilt smearing are considered in our calculations. The optimization leads to small changes ($<$ 1.0 \%) in the lattice constants. These spin-polarized calculations start with aligning all eight Fe atoms ferromagnetically with large moments while the one Mn atom is aligned antiferromagnetically to the Fe atoms in the total 16--atom unit cell. However, after the convergence, the negligibly small Fe and large Mn magnetic moments are aligned ferromagnetically. We discuss these results in connection with the magnetic properties of the compositions at a later stage. 

\begin{figure}[ht]
 \includegraphics[width=\linewidth]{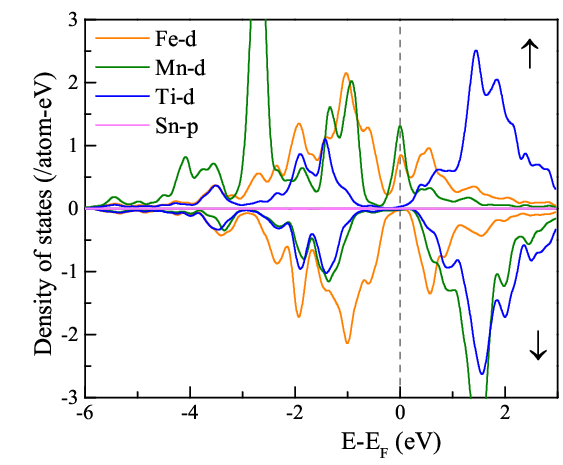}
\caption{Atom-projected spin-polarized density of states for Fe, Mn, and Ti -$d$ and Sn $p$ orbitals in Fe$_2$Ti$_{0.75}$Mn$_{0.25}$Sn in the ferromagnetic arrangements. Contributions from Fe and Mn $d$ states in the up spin channel at the Fermi level (set to zero) signify the itinerant magnetic interaction between these two atoms. Contributions from Ti $d$ and Sn $p$ orbitals are either below or above the Fermi level.}
    \label{fig: Fig.2}
    \end{figure}

For the present calculation, the Mn-Mn magnetic interaction is estimated by considering 2$\times$2$\times$2 supercell of 128 atoms and replacing two Ti atoms with Mn. The different separation between these two Mn atoms within this supercell is then considered to extract the distance-dependent magnetic interaction. The four possible Mn-Mn separations considered, after avoiding the possible interference of periodic images of supercell on magnetic interactions, are 4.254, 6.017, 7.369, and 8.509 \AA{}. At these four distances, magnetic interaction is estimated by calculating the energy difference between ferromagnetic and antiferromagnetic arrangements of the two Mn atoms. Background from negligibly small Fe and Ti magnetic moments has an insignificant influence on the magnetic interactions estimated in this fashion.

\section{Results and Discussion} 

 \begin{figure}[ht]
    \centering
     \includegraphics[width=\linewidth]{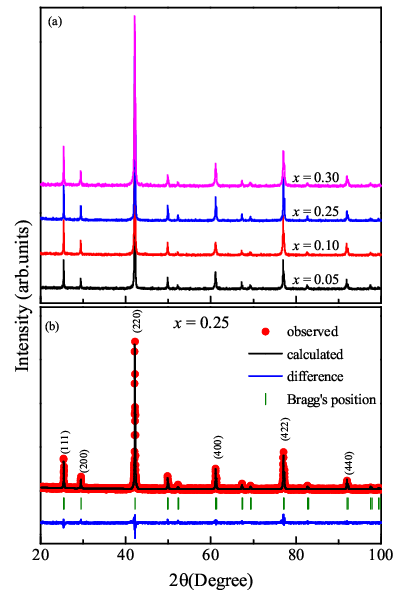}
   \caption{(a) X-ray diffraction profile of Fe$_2$Ti$_{1-x}$Mn$_x$Sn, and (b) Rietveld refined XRD profile of $x$ = 0.25.}
    \label{fig: Fig.3}
\end{figure}

\begin{figure*}[ht]
 \includegraphics[width=0.8\textwidth]{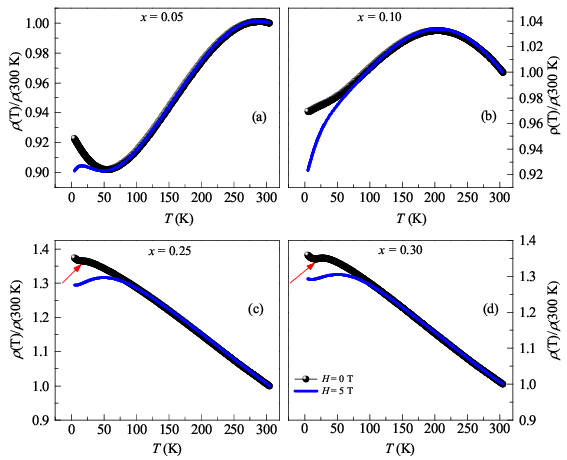}
 \caption{Temperature dependent electrical resistivity of Fe$_2$Ti$_{1-x}$Mn$_x$Sn measured at an applied magnetic field of 0 T and 5 T.}
    \label{fig: Fig.4}
    \end{figure*}
    
The density of states (DOS) scenario of the Fe$_2$Ti$_{1-x}$Mn$_x$Sn system is obtained using a DFT-only approach. For this case, the atom-projected spin-polarized DOS of Fe, Mn, Ti -$d$ and Sn $p$ states are shown in Fig.\ref{fig: Fig.2}. Many insightful observations can be made from this plot.  First, the system shows the half-metallic character of the DOS with mainly Fe and Mn $d$ states in the spin-up channel, while there is a band gap in the spin-down channel at the Fermi level. Though Ti $d$ states also contribute below the Fermi level, the dominant contributors remain the $d$ states of Fe and Mn atoms, and the magnetic properties arise from the interactions among all three transition metal atoms. Second, this interaction can be considered itinerant as the system is metallic. Third, the magnetic moment on the Mn site is large, while Fe and Ti atoms carry negligible moments, and both Mn \& Fe have finite contributions to the DOS at the Fermi level. Recall that we started with a scenario of antiferromagnetic coupling between single Mn and eight (ferromagnetically aligned) Fe atoms. Such a system converges to a ferromagnetic coupling among all the magnetic atoms, with negligibly small magnetic moments on Fe ($\sim$ 0.016 $\mu_B$) and Ti ($\sim$ 0.052 $\mu_B$) atoms and a relatively high moment on Mn ($\sim$ 3.161 $\mu_B$) atoms. It appears as if Mn atoms are the ones with localized moments that are interacting through the cloud of conduction electrons formed by Fe and Ti atoms. 

With the intent of finding a half-metallic Heusler, we begin the experimental investigation into the Mn--substituted Fe$_2$TiSn compounds. The X-ray diffraction (XRD) patterns of Fe$_2$Ti$_{1-x}$Mn$_x$Sn ($x$ = 0.05, 0.1, 0.25, and 0.3) are shown in Fig.\ref{fig: Fig.3}(a). The presence of strong (111), (200), and (220) peaks confirm the highly ordered cubic L2$_1$ crystal structure in all the prepared compositions. However, the solubility limit of Mn in Fe$_2$TiSn is limited to 30$\%$ as additional peaks related to the secondary phase begin to appear for compositions with $x\geq$ 0.4. Contrary to the general expectation of finding an increase in the antisite disorder upon carrying out substitutions in an intermetallic system, the present compositions show a relative increase in the intensity of the (111) and (200) super-lattice peaks, indicating a reduction of the overall disorder in comparison to pristine Fe$_2$TiSn. Rietveld refinement of the XRD profiles, implemented through the FullProf suite\cite{FullProf}, yield values for lattice parameters that follow Vegard's law\cite{Vegard}, indicating the desired substitutions at the targeted lattice sites have indeed been achieved. 

With the prediction of half-metallic DOS at the Fermi level and a highly ordered crystal structure, studying the electrical resistivity of the Mn--substituted Fe$_2$TiSn compounds becomes pertinent. It has been previously observed that the electrical resistivity, $\rho(T)$, of Fe$_2$TiSn, shows a bad metallic behavior with a low-temperature upturn and minimum at 50 K. This low-temperature upturn in $\rho(T)$ is due to weak localization caused by anti-site disorder between the Fe and Ti sites. Fig.\ref{fig: Fig.4} shows the $\rho(T)$ data for Fe$_2$Ti$_{1-x}$Mn$_x$Sn ($x$ = 0.05, 0.10, 0.25, 0.30) for a temperature range of 5--300 K and at applied magnetic fields of 0 T and 5 T. For brevity, each composition's $\rho(T)$ is normalized to $\rho(300 K)$ value. The general features of $\rho(T)$ curve of $x$ = 0.05 sample match that of Fe$_2$TiSn\cite{sayanJPCM1}. However, when viewed from the high-temperature side of the $\rho(T)$ curve for $x$ = 0.10 sample shows a negative temperature coefficient of resistance (TCR) down to 200 K, followed by positive TCR at low temperatures. On the other hand, $x$ = 0.25 and 0.30 compositions exhibit complete negative TCR throughout the measurement range. A closer inspection reveals that the $\rho(T)$ of $x$ = 0.25 and 0.30 shows a weak slope change at low temperature (indicated by arrow), below which $\rho(T)$ increases at a faster rate with decreasing temperature.

\begin{table*}[ht]
  \begin{center}
    \caption{ The extracted parameters after fitting the $\rho$(T) data in low and high-temperature regions for the applied fields of 0 and 5T.}
    \label{tab: Table_2}
    \renewcommand{\arraystretch}{1.5}
    \setlength{\tabcolsep}{10pt}
    \begin{tabular}{c c c c c c c c c c c}
    \hline
     $x$ & \multicolumn{2}{c}{$\rho_0$} & \multicolumn{2}{c}{$\rho_{e}$$\times$10$^{-2}$} &  \multicolumn{2}{c}{$\rho_K$$\times$10$^{-2}$}  & \multicolumn{2}{c}{$\rho_p$$\times$10$^{-9}$} &  $\rho_{ep}$$\times$10$^{-3}$ & $\rho_{em}$$\times$10$^{-7}$\\
        & 0T & 5T & 0T & 5T & 0T & 5T & 0T & 5T &  \\ 
      \hline
     0.25&1.389&1.292&0.67&1.20&1.89&1.7&1.239&0.73&-1.18&-5.909\\
     0.30&1.376&1.297&1.01&1.24&2.5&1.9&0.397&0.135&-1.29&-2.42\\
     \hline
    \end{tabular}
  \end{center}
  \end{table*}

\begin{figure*}[ht]
\centering
\includegraphics[width=0.75\textwidth]{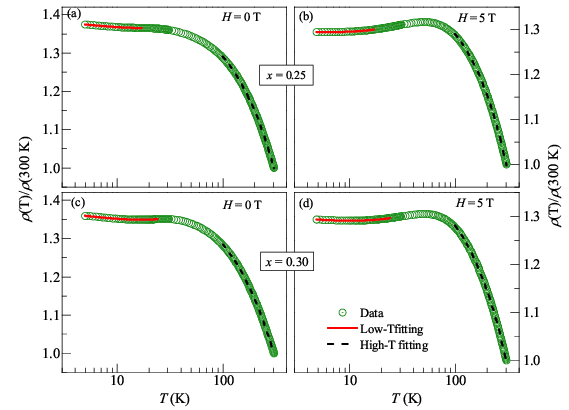}
 \caption{(Color online) (a)-(d) shows the fitting of temperature-dependent resistivity data of $x$ = 0.25 and 0.30 at 0 T and 5 T, respectively. The temperature axis is represented in the log scale for better visibility of the low-$T$ fitting.}
    \label{fig: Fig.5}
\end{figure*}

The low-temperature data is analyzed by considering the different scattering mechanisms that may contribute to the observed $\rho(T)$ behavior. A low-$T$ upturn in resistivity can result from various mechanisms, including electron-electron interaction (EEI), weak localization (WL), and Kondo-like transport. EEI arises from electron-electron elastic scattering due to the Coulomb interaction between conduction electrons, WL arises from disorder-mediated coherent backscattering of charge carriers, and Kondo-like transport arises from the interaction between localized magnetic moment associated with the magnetic impurities and mobile conduction electrons\cite{TV_ramakrishnan, KFM_2}. In Mn substituted Fe$_2$TiSn, a reduction in atomic disorder and/or magnetic order destroys the WL, suggesting that the low-T upturn of $\rho(T)$ in compositions with higher Mn concentration is not caused by WL\cite{sayanJPCM1, sayan_JAP}. Furthermore, we fit the $\rho(T)$ data separately with EEI and Kondo effect. As the result indicates, neither model fits the experimental data satisfactorily. Thus, combining both mechanisms likely contributes to the observed low-T upturn in resistivity. Besides, the contribution from electron-phonon scattering should also be added to the form $T^5$ to represent the resistivity\cite{Chatterjee_2022}. Thus, the low-T resistivity data at H = 0 T and 5 T can be described by the expression $$\rho(T) = \rho_0 + \rho_{e}T^{1/2} -\rho_{K}ln(T) +\rho_{p}T^5$$, where, $\rho_0$ is residual resistivity, and, $\rho_e$, $\rho_K$ and $\rho_p$ represent coefficients for EEI, Kondo-like transport and electron--phonon scattering respectively. The fitting result is shown in Fig.\ref{fig: Fig.5}, and the fitting parameters for different Mn substitution and applied magnetic field are depicted in Table-\ref{tab: Table_2}. Both $\rho_e$ and $\rho_K$ increase with increasing $x$. The value of $\rho_K$ for a given composition is higher at 0 T compared to 5 T, highlighting the significant role of magnetic scattering of charge carriers. Whereas not only the value of $\rho_e$ is higher at 5 T as compared to that of 0 T, but it also has a positive value. However, if the effect of both Kondo and EEI are simultaneously taken into account, $\rho_e$ may end up being positive\cite{KFM_1}. Generally, when EEI alone contributes to the $\rho(T)$, the $\rho_e$ value extracted from fitting the data yields a negative value.

 Next, we analyze the behavior of $\rho(T)$ of $x$ = 0.25 ( and 0.30) composition in the high-T region. While the negative TCR has been observed in Heusler compounds\cite{Co_Fe_Ga, SG_Mn2CoAl, LuNiSb, CrVTiAl, CoFeCrGa}, the notable feature here is the linear $\rho(T)$ between 100 K and 300 K, which is unusual compared to the common semiconductor and disordered metallic systems. In semiconductors, electron transport exhibits a thermally activated behavior of $\rho(T)$ data where $\rho$ vs T plot is upward convex. On the contrary, disordered metallic systems show a convex downward behavior of $\rho(T)$ data. Further, reports on the high-T resistivity data of Fe$_2$TiSn show a negative TCR from 300 K to 800 K, which is attributed to the interband transition of conduction electrons through a gap at Fermi level\cite{Slebarski_2006}. With electron doping in pristine Fe$_2$TiSn, this negative TCR persists below 300 K\cite{FTS_kondo} and may also result in half-metallicity\cite{sayanJPCM2,sayanSR}. A similar signature is observed in the present Fe$_2$Ti$_{1-x}$Mn$_x$Sn compositions. In the case of half-metals, the high-temperature electrical resistivity can be expressed by the combined contribution from electron-magnon and electron-phonon scattering. The contribution of electron-magnon scattering towards resistivity follows T$^2$ dependence and the contribution of the scattering of conduction electron by phonons is linear-T at high temperatures. Thus, we have fitted the resistivity data in the high-T region using the equation: $$\rho(T) = \rho_0 + \rho_{ep}T +\rho_{em}T^2$$, where $\rho_{ep}$ and $\rho_{em}$ represent the coefficients of electron-phonon and electron-magnon scattering, respectively. The fitting results are shown in Fig.\ref{fig: Fig.5}. Two observations can be made from the values of the $\rho_{ep}$ and $\rho_{em}$ (given in Table-\ref{tab: Table_2}), both the coefficients are negative and the magnitude of $\rho_{em}$ is very small compared to the $\rho_{ep}$. The negative value of the coefficients is expected as $\rho(T)$ increases with a decrease in temperature\cite{Co_Fe_Ga}. Additionally, since electron-magnon scattering is a spin-flip process, the lower value of $\rho_{em}$ as compared to the $\rho_{ep}$ denotes the presence of spin-polarised DOS at Fermi level in $x$ = 0.25 (and 0.30). Our theoretical calculations for $x$ = 0.25 also support this observation. Thus, the combined results of theoretical calculations and electrical resistivity indicate the presence of half metallicity in Mn-substituted Fe$_2$TiSn.

\begin{figure}[ht]
\centering
\includegraphics[width=1\linewidth]{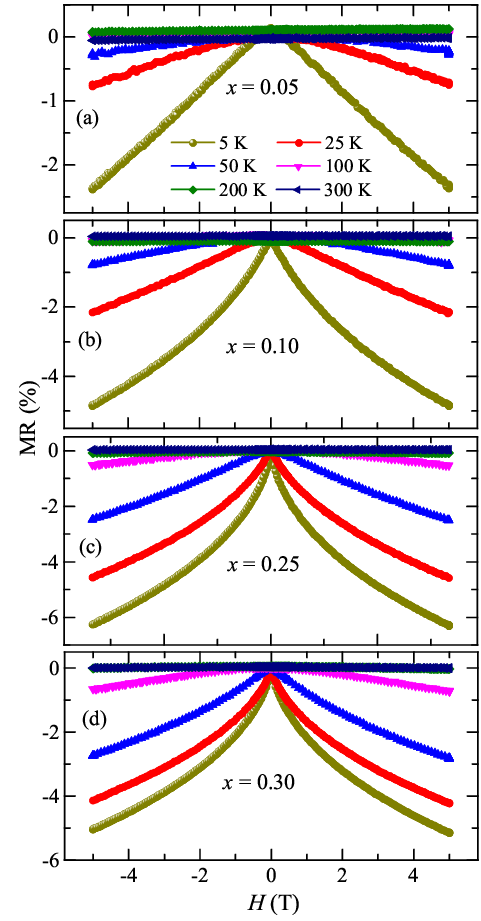}
 \caption{Field dependent isothermal magnetoresistance, $MR$, of Fe$_2$Ti$_{1-x}$Mn$_x$Sn.}
    \label{fig: Fig.6}
\end{figure}

Alternatively, the magnetic field dependence of electrical resistance at fixed temperatures also presents the same scenario as that of $\rho$(T). Fig.$\ref{fig: Fig.6}$ shows the isothermal magnetoresistance (MR) curves measured at different temperatures ranging from 5 K$\leq$T$\leq$300 K and for all four compositions. Using the standard definition of MR as, $MR(\%)=\left[\frac{\rho(H) - \rho(0)}{\rho(0)}\right]$$\times$100, we find that all the Mn--substituted compositions show the isotropic negative MR in the entire temperature range. Generally, the field-dependent resistivity, $\rho(T,H)$, depends on the combined sum of Lorentz force and contribution from different scattering mechanisms. The Lorentz force constitutes the classical MR, which is inherently positive. Meanwhile, in the studied compositions, the high-T electrical resistivity is mainly governed by electron-phonon and electron-magnon scattering, where the former is nearly independent of the applied magnetic field. The combined effect of these two processes likely leads to the observed small yet negative MR at high temperatures\cite{MR_2}.

Moreover, the increase in the magnitude of the low-T MR with increasing Mn content suggests the presence of magnetic ordering and/or spin-dependent scattering. Additionally, compositions showing signatures of half-metallicity, i.e. $x$ = 0.25 (and 0.30), exhibit the dominance of EEI and the Kondo effect in low-T resistivity. Notably, the EEI effect leads to a positive MR due to Zeeman splitting and orbital effects, with its magnitude varying as the square root of the applied magnetic field\cite{TV_ramakrishnan, MR_TVR, MR_1, MR_2, MR_4}. This has been reflected as the increment of $\rho_e$ value with an applied magnetic field of 5 T for a given value of $x$(see Table-\ref{tab: Table_2}). Conversely, applying the magnetic field suppresses the fluctuation in the localized magnetic moment and spin-dependent scattering, leading to a negative MR in a Kondo system\cite{MR_3}. Thus, for $x$ = 0.25 (and 0.30), the negative MR supports that the Kondo effect is the dominant mechanism governing the low-T electrical resistivity. This also explains the higher value of $\rho_k$ as compared to the $\rho_e$ obtained from the fitting of the low-T resistivity data.


\begin{figure}[ht]
    \centering
     \includegraphics[width=\linewidth]{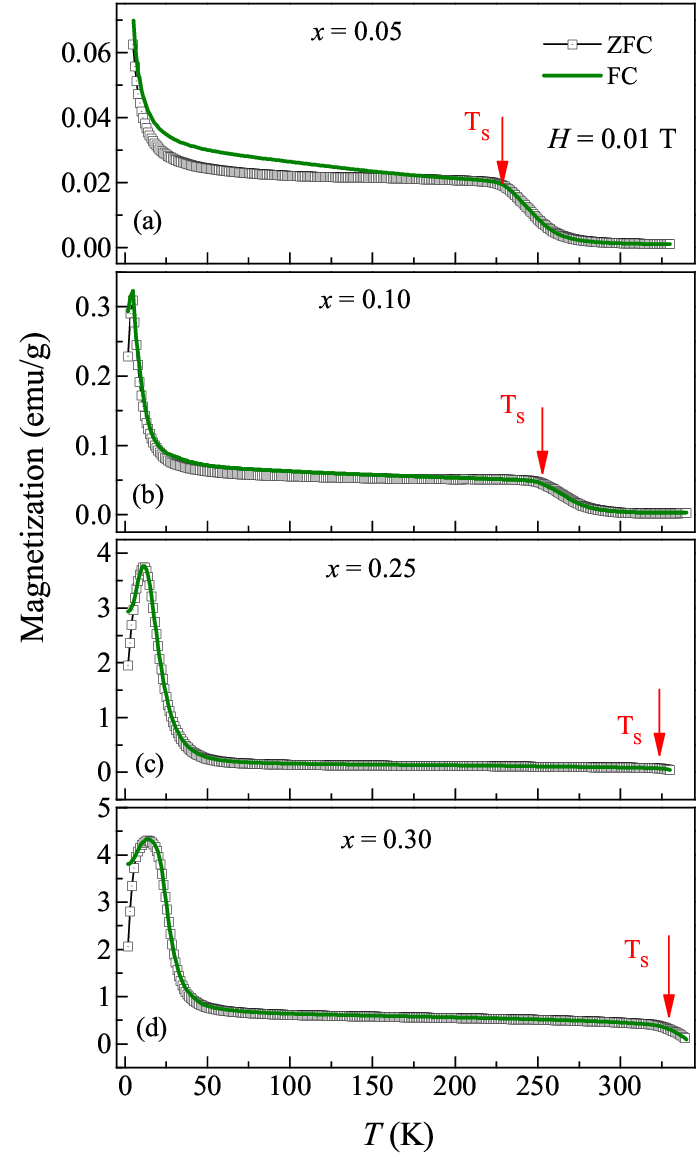}
   \caption{Temperature dependence of the magnetization, $M(T)$ curves, of (a) $x$ = 0.05, (b) $x$ = 0.10, (c) $x$ = 0.25, and (d) $x$ = 0.30.}
    \label{fig: Fig.7}
\end{figure}

\begin{figure}[ht]
    \centering
     \includegraphics[width=\linewidth]{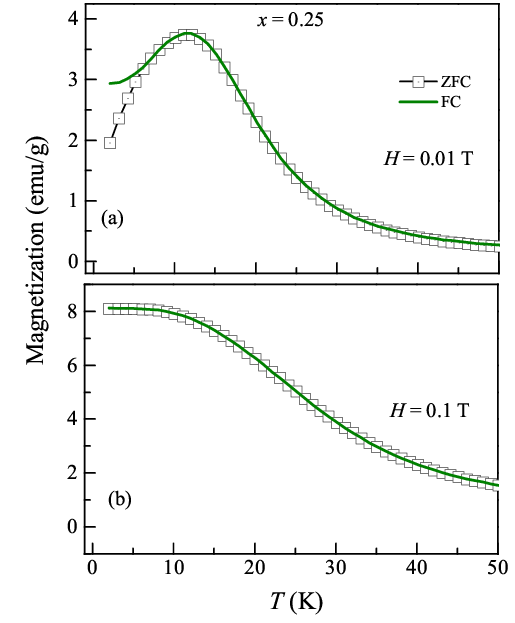}
   \caption{Temperature dependence of the magnetization, $M(T)$ curves, of $x$ = 0.25 at an applied magnetic field of (a) $H$ = 0.01 T, and (b) $H$ = 0.1 T.}
    \label{fig: Fig.8}
\end{figure}

Having understood the electronic DOS in the Fe$_2$Ti$_{1-x}$Mn$_x$Sn compositions and half-metallic ground state in $x$ $\geq$ 0.25, we now discuss the experimental study of its magnetic properties. The temperature-dependent magnetization, $M(T)$, measured at 0.01 T, is plotted in Fig.\ref{fig: Fig.7}. According to the previous studies, an inherent Fe/Ti antisite disorder in the otherwise non-magnetic Fe$_2$TiSn induces the formation of weak magnetic clusters\cite{sayanJPCM1}. These interactions are identified by a step-like feature in the high-temperature region, denoted here as T$_S$, in the M(T) curve. This step-like feature shifts towards the high-temperature side with rising Mn substitution. Additionally, an abrupt rise in magnetization is observed at low-- temperatures, marking the transition to a magnetically ordered state. This temperature is identified as T$_C$, and its value increases with rising Mn content. The values of T$_C$ and T$_S$ are listed in the Table-\ref{tab: Table_1}.

For small values of applied magnetic field, there is a noticeable contrast between the M(T) plots recorded during the zero field cooled (ZFC) and field cooled (FC) protocols. This bifurcation between ZFC/ FC disappears when a relatively strong magnetic field of $\geq$ 0.1 T is applied (Fig.$\ref{fig: Fig.8}$), indicating the presence of weak anisotropy in the material. In the ZFC process, the material's anisotropy determines the amount of magnetization below T$_C$ and may prevent a relatively small applied field from aligning the moments in the direction of the field. When the same material is cooled through T$_C$ while being exposed to a magnetic field during the FC process, the moments are locked into the direction of the applied magnetic field. As a result, there is a noticeable contrast between the ZFC and FC curves for small applied fields, and this bifurcation disappears when a slightly stronger magnetic field is applied.

\begin{figure}[ht]
    \centering
     \includegraphics[width=\linewidth]{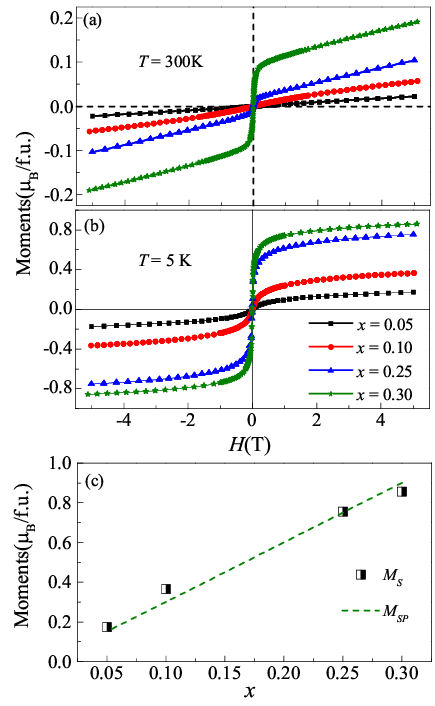}
   \caption{Isothermal magnetization, $M(H)$ curve of all the compositions measured at (a) 300 K, (b) 5 K. (c) shows the values of experimental saturation magnetization ($M_S$) at 5 K, plotted along with the Slater-Pauling rule (dotted line) for varying Mn concentration. }
    \label{fig: Fig.9}
\end{figure}

Fig.\ref{fig: Fig.9}(a-b) shows the variation of isothermal magnetization ($M$) as a function of the applied magnetic field ($H$) for all the compositions measured at 300 K and 5 K, respectively. For $x$ = 0.05, the $M(H)$ curve at 300 K displays a linear behavior (Fig.\ref{fig: Fig.9}(a)). Further increase in $x$, a sharp rise in $M$ at lower fields is observed followed by a linear component at higher fields. These results indicate the presence of weak ferromagnetic clusters at 300 K. Consequently, the $M(H)$ curve measured at $T$ = 5 K shows characteristics of a ferromagnetic ordering with an increase in the magnetic moment values with $x$ (see Fig.$\ref{fig: Fig.9}$(b)). The saturation magnetic moment per formula unit is found to follow the Slater-Pauling rule (Fig. \ref{fig: Fig.9}(c)), reflecting possibly the spin-polarized band structure of Fe$_2$Ti$_{1-x}$Mn$_x$Sn compositions. However, it may be noted that the $M(H)$ curves do not reach complete saturation till $H$ = 5 T. The presence of ferromagnetic clusters at $T$ = 300 K suggests that the low-temperature magnetization behavior could be due to the evolution of these clusters. To interpret this, we have fitted the $M(H)$ curve using the modified Langevin function, expressed as $M(H) = M_s L(\alpha) + \chi H$, where $\alpha$ = $\mu$H/k$_B$T, M$_S$ represents the saturation magnetization, $\mu$ is the average magnetic moment and $\chi$ represents the paramagnetic susceptibility. The lack of fit of this model to the data (not shown here) rules out any cluster-like evolution in the prepared compositions.

Alternatively, from the measurements, it appears that the Mn-substituted Fe$_2$TiSn compositions display a long-range ferromagnetic order below $T_C$, which exhibits a substantial dependency on the strength of the applied magnetic field with respect to the random anisotropy. Further, the $\rho(T)$ data shows the Kondo-like behavior, i.e., the presence of local Mn moment screened by conduction electrons, a weak random anisotropy model\cite{RAM} could be applicable here. Such a model has previously been extensively used to explain the short-range FM correlations in Heusler alloys\cite{RAM_ex1, RAM_ex2, RAM_ex3}. This model incorporates both the experimentally applied magnetic field and the ground state configuration of magnetic materials with random anisotropy for a wide range of anisotropy strengths. Three distinct regimes can be identified for the weak anisotropy situation based on the magnitude of the applied magnetic field, H, and the parameter $H_S$ (= $H^4$$_r$/$H^3$$_{ex}$), where $H_r$ and $H_{ex}$ are anisotropic and exchange fields, respectively. For $H$ $<$ $H_S$, one obtains a correlated spin glass phase with a high magnetic susceptibility for low fields. When $H_S$ $<$ $H$ $<$ $H_{ex}$, a ferromagnet with a wandering axis (FMWA) results from a rough alignment of the spins in the presence of a weak magnetic field. The tipping angle of the magnetization vector, with respect to the applied magnetic field, changes throughout the system due to the non-collinear magnetic structure. In this regime, the magnetization approaches saturation as,

\begin{equation}
    M(H)=M_s \left[1 -\frac{1}{15} \left(\frac{H_s}{H + H_c}\right)^{\frac{1}{2}}\right]
\end{equation}

where $M_s$ is saturation magnetization, and $H_c$ is the field due to the coherent portion of the anisotropy. In the third regime where $H$ $\gg$ $H_S$, the correlation length of the spins reduces as the field strength increases, whereas the tipping angle is completely uncorrelated amongst sites.  Except for a slight tipping angle (less than in the case of FMWA) brought on by random anisotropy, all spins in this regime are almost aligned with the field. In such cases, isothermal magnetization gradually approaches saturation as,

 \begin{equation}
     M(H)=M_s \left[1 -\frac{1}{15}\left(\frac{H_r}{H+H_{ex}}\right)^2\right]
 \end{equation}

\begin{figure}[ht]
    \centering
     \includegraphics[width=\linewidth]{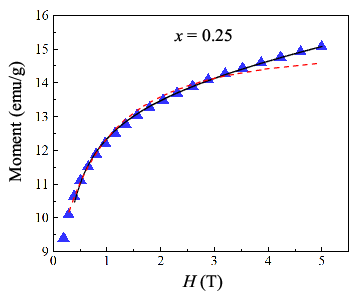}
   \caption{ (Color online) Fitting of $M(H)$ curve at $T$ = 5 K with random anisotropy model for $x$ = 0.25. The solid line shows the fitting with equation (1), and the dashed line shows the fitting to equation (2). See the text for more details.}
    \label{fig: Fig.10}
\end{figure}

We have fitted equation(1) with an additional paramagnetic contribution,$\chi$H\cite{RAM_fit1}, to the M(H) curve at 5 K of $x$ = 0.25, as shown by the solid line in Fig.\ref{fig: Fig.10}. It shows a satisfactory fit for a wide range of applied fields from 0.4 to 5 T. The value of H$_s$ derived after the fitting ranges from 0.303 T for $x$ = 0.05 to 0.021 T for $x$ = 0.30, which is lower than the maximum applied field, following the condition for the system to fall in the second regime. In contrast, equation(2) does not yield a good fit to the data (see the dashed line in Fig.\ref{fig: Fig.10}), and the value of $H_s$ derived after the fitting is much higher than the maximum applied field despite using the prefactor equal to 1 instead of 1/15\cite{RAM_ex2}.
 
\begin{figure}[ht]
    \centering
     \includegraphics[width=\linewidth]{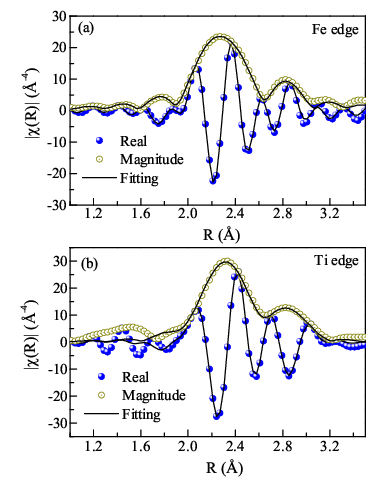}
   \caption{Magnitude (hollow spheres) and real component (solid spheres) of Fourier-transformed (a) Fe, and (b) Ti K edge EXAFS spectra at 6 K for $x$ = 0.25. See the text for details.}
    \label{fig: Fig.11}
\end{figure}

Random anisotropy in cubic systems can occur from substitutional impurities and disorder\cite{Cubic_RAM}. In Ref.\cite{RAM_fit1}, P.M. Gehring \textit{et al.} proposes that the random anisotropy in cubic polycrystalline Dy$_{1-x}$Y$_x$Al$_2$ could originate from lowering of local lattice symmetry, driven by lattice mismatch, difference in charge screening and spin-orbit interactions\cite{RAM_fit1}. In certain cubic Heusler systems, such random anisotropy is found to influence its low-temperature magnetic properties \cite{RAM_ex1}. To unravel the reason that the random anisotropy in our cubic Fe$_2$Ti$_{1-x}$Mn$_x$Sn originates from local lattice distortion, we undertake measurement of the high-resolution synchrotron-based EXAFS spectra at Fe and Ti K-edges for $x$ = 0.25. The EXAFS data is extracted in the reciprocal space, $\chi(k)$, and Fourier transformed to real space, $\chi(R)$, where the magnitude and real part of $\chi(R)$ reflects the radial distribution function around the absorbing atoms. The local structural refinement is carried out by varying the atomic distances and thermal mean square factor ($\sigma^2$) until a good fit is obtained in the range of 1.8--3.0 \AA~ in the $R$ space and 3 to 14 \AA$^{-1}$ in $k$ space. Based on the information extracted from the XRD profile of the cubic L2$_1$ structure and lattice constant of 6.059 \AA~, the local atomic arrangement around the Fe atom has 4 Ti and 4 Sn atoms at an equal distance of $\sim$ 2.623 \AA, and 6 Fe atoms at $\sim$ 3.029 \AA. Similarly, for the Ti atom, there are 8 Fe atoms at $\sim$ 2.623 \AA~ and 6 Sn atoms at $\sim$ 3.029 \AA. Given its cubic symmetry, all these bond distances can be represented as a multiple of lattice constant, and, in principle, only varying the lattice constant should give a good fit. However, this model does not fit the experimental data satisfactorily. Whereas a better fit is achieved by independently varying the individual bond distances, as shown in Fig.\ref{fig: Fig.11}, and the obtained fitting parameters are listed in Table-\ref{tab: Table_3}. The notable aspect is the discrepancy in the Fe--Ti bond distance when viewed separately from the Fe and Ti edges. From the Ti-edge, the Ti--Fe bond distance comes out to be $\sim$2.613 \AA, consistent with the value derived from the lattice constant, whereas the same bond distance, when viewed from the Fe edge, yields a value of 2.596 \AA. Furthermore, the Fe--Sn bond distance observed from the Fe-edge is 2.621 \AA, which aligns well with the lattice constant. This observation contradicts the expectations of the cubic symmetry where Fe--Ti and Fe--Sn bond distances should be equidistant from the Fe atom. This discrepancy in the Fe--Ti correlation comes from the fact that Mn replaces some Ti atoms in Fe$_2$Ti$_{1-x}$Mn$_x$Sn compositions, and the bond distance involving the Ti site represents an average distance of Fe--Ti and Fe--Mn. This further indicates that in the Fe--(Ti/Mn) correlation, where Mn occupies the vacant Ti sites, the local lattice environment is distorted, leading to variations in the bond distance. These observations confirm that it is the lowering of local symmetry which causes the random anisotropy in Fe$_2$Ti$_{1-x}$Mn$_x$Sn compositions.

\begin{table}[t]
  \begin{center}
    \caption{The coordination number (CN), bond distance (R), and thermal mean square variation ($\sigma$$^2$) from EXAFS data fitting. $R_{eff}$ represents the expected bond distance calculated from the experimental lattice constant.} 
    \label{tab: Table_3}
    \setlength{\tabcolsep}{5pt}
    \renewcommand{\arraystretch}{1.5}
     \begin{tabular}{l c c c c r}\\
     \hline
   Bond type & CN  & $R_{eff}$ & R(\AA) & $\sigma$$^2$(\AA$^{2}$) \\
        \hline
        &   &  Ti K-edge &  &   \\
        \hline
      Ti-Fe & 8 & 2.623  &2.613(5) & 0.00376(38) \\
     Ti-Sn & 6 & 3.029 &3.023(9) & 0.00543(76)  \\
    \hline 
      &   &  Fe K-edge &  &   \\
        \hline
      Fe-Sn & 4 & 2.623  & 2.621(1) & 0.00307(15)  \\
    Fe-Ti/Mn & 4 & 2.623 & 2.596(3) & 0.00568(45)   \\
   Fe-Fe & 6 & 3.029  & 3.008(4) & 0.00767(43)  \\
    \hline
    \end{tabular}
  \end{center}
 \end{table}

\begin{figure}[ht]
 \includegraphics[width=\linewidth]{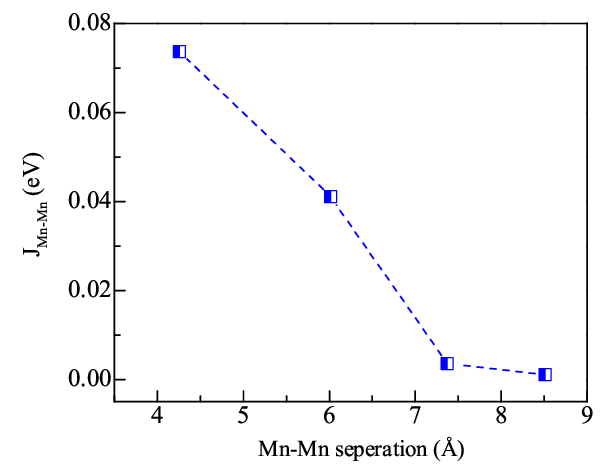}
\caption{Mn-Mn magnetic interaction at various separations in the unit cell of $x$ = 0.25. For details, refer to the "Experimental and computational details" section.}
    \label{fig: Fig.12}
\end{figure}

Thus, the analysis of \textit{ab initio} DOS calculation, resistivity, and magnetic data suggest that Mn substitution in Fe$_2$TiSn introduces localized moments. These local moments exhibit strong coupling with the conduction electron, which screens the interaction between two Mn atoms. As the results of this screening, the strength of the magnetic interaction decreases with increasing Mn-Mn distance, and typical ferromagnetic order is absent in Fe$_2$Ti$_{1-x}$Mn$_x$Sn. We have calculated the distance-dependent magnetic exchange interaction (J$_{Mn-Mn}$) between two Mn atoms to verify this. Estimating distance-dependent Mn-Mn magnetic interaction with smaller unit cells (considered for 25$\%$ substitution) is difficult. This is because, at large Mn-Mn separation in a small unit cell, contributions from the periodic images of Mn atoms at relatively shorter distances will start contributing to the magnetic energy. To avoid this problem, one has to consider a large supercell. We consider a 2$\times$2$\times$2 supercell and replace two Ti with Mn atoms. The distance-dependent Mn-Mn magnetic interaction (J$_{Mn-Mn}$ = E$_{AFM}$-E$_{FM}$) is then estimated as explained in the "Experimental and Computational details" section. Here, we take advantage of the fact that the magnetic moments on Fe and Ti atoms are negligibly small, and hence, their insignificant direct contributions to J$_{Mn-Mn}$ can safely be ignored. The plot of separation dependent J$_{Mn-Mn}$ is shown in Fig.~\ref{fig: Fig.12}. One can observe that the effective Mn-Mn interaction is ferromagnetic, consistent with the experimental findings. This ferromagnetic interaction exponentially decreases with the Mn-Mn separation and becomes negligible at the separation of 8.51\AA.

\begin{figure}[ht]
 \includegraphics[width=\linewidth]{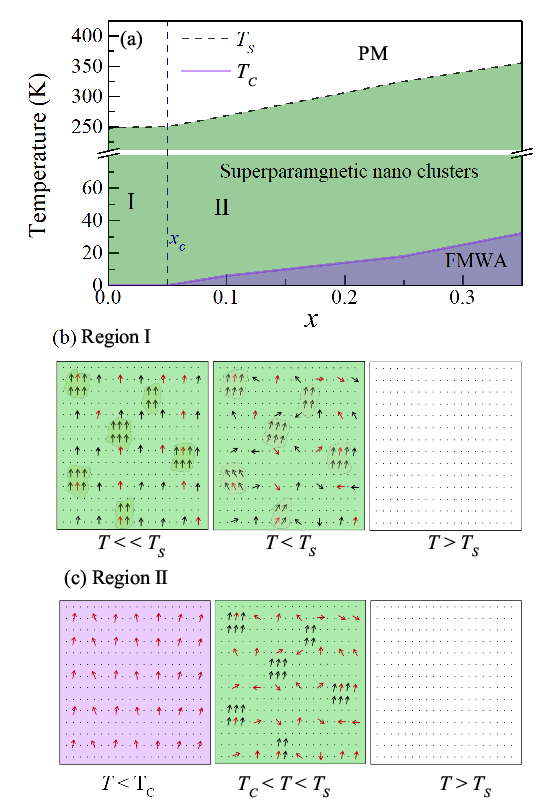}
\caption{$x$-T phase diagram representing the change in the magnetic state of Fe$_2$Ti$_{1-x}$Mn$_x$Sn with the decrease in the temperature. The local Mn moments (red arrow) align at an angle with the direction of the applied magnetic field.}
    \label{fig: Fig.13}
\end{figure}

Hence, according to the results of magnetic measurements, the magnetic state of Fe$_2$Ti$_{1-x}$Mn$_x$Sn changes with temperature as marked in the $x$-T phase diagram, as shown in the Fig.\ref{fig: Fig.13}(a). Fe$_2$Ti$_{1-x}$Mn$_x$Sn is in a typical PM state when the temperature is higher than T$_s$. As temperature decreases, ferromagnetic correlation appears due to the formation of magnetic clusters. In the case of $x$ $<$ $x_c$ (region-I), clusters continue to evolve with falling temperatures and lead to a superparamagnetic ground state, and no long-range magnetic order is observed in this region. The simplified schematic diagram of the magnetic state as a function of temperature is presented in Fig.\ref{fig: Fig.13}(b-c). Substituting higher Mn content ($x$ $>$ $x_c$) in place of Ti in Fe$_2$Ti$_{1-x}$Mn$_x$Sn significantly alters this magnetic behavior. The strong local moments of Mn suppress the influence of high-temperature FM clusters on the low-temperature magnetic ground state and result in an unconventional ferromagnetic ordering driven by random anisotropy, known as the "ferromagnet with wandering axis" state, as represented as region-II in the phase diagram. This transition from a superparamagnetic state in Fe$_2$TiSn to a weak random anisotropy-driven state with Mn substitution underscores the significant impact of Mn substitution on the magnetic properties of Heusler alloys. Furthermore, as confirmed from electrical resistivity and magnetic measurements and further supported through our DFT calculations, Mn substitution induces half-metallicity in the Fe$_2$Ti$_{1-x}$Mn$_x$Sn, enhancing its potential for spintronic applications.

It is crucial to highlight that analysis of low-temperature resistivity and isothermal magnetization presents two seemingly divergent scenarios: the Kondo effect and ferromagnetic order and coexist in the same system. The Kondo effect is generally suppressed in a strong ferromagnet characterized by high interaction strength and ordering temperature because magnetic ordering fixes the local moment direction. However, there have been reports of simultaneous observation of Kondo scattering and ferromagnetism in heavy fermion systems with partially filled 4f or 5f shells\cite{KFM_1, KFM_2, KFM_3, KFM_4}, which are theoretically modeled as a regular lattice of local moments coupled to conduction electrons, and described as underscreened Kondo lattice model\cite{KFM_model}. Besides, this observation is not limited to the heavy fermionic system with f-spins but also observed in simple 3d electron systems. For instance, Pasupathy et al. have observed Kondo-assisted tunneling via C$_{60}$ molecules in contact with ferromagnetic nickel electrodes\cite{KFM_science}. Additionally, thin films of TiO$_2$ and Pt-substituted MnBi have shown the coexistence of ferromagnetism and Kondo effect\cite{KFM_TiO2, KFM_MnBi}. Recently, the ferromagnetism-induced Kondo effect has been observed in magnetically doped graphene, and this doped ferromagnetic graphene behaves as half-metal\cite{KFM_graphene}. Moreover, Kondo-dominated low-temperature resistivity is also observed in Fe$_{2-x}$Ni$_x$TiSn Heusler system\cite{FTS_kondo}. 

\section{Conclusion}
This study investigated structural, magnetic, and transport properties of Mn substituted Fe$_2$TiSn through room temperature XRD, temperature and field-dependent magnetization, and magnetotransport measurements. Room temperature XRD patterns confirm the single phase formation of all the compositions with order L2$_1$ structure. The temperature variation of electrical resistivity Fe$_2$Ti$_{1-x}$Mn$_x$Sn shows an overall increase in the magnitude with increasing Mn content.
For $x\geq$ 0.25, $\rho(T)$ shows a negative temperature coefficient of resistance, followed by an upturn at $T$ $<$ 20 K due to Kondo-like scattering and EEI. High temperature $\rho(T)$ is dominated by electron-phonon scattering,  indicating spin-polarized states. Electrical transport and the density of states calculation confirm the formation of highly spin-polarized electronic structures and the presence of half-metallicity on Mn substituted Fe$_2$TiSn. Magnetization measurements display features indicative of weak anisotropy in the system, and low-temperature $M(H)$ behavior aligns with the random anisotropy model, suggesting that a ferromagnet with a wandering axis can be an apt description of the magnetic ground state of the Fe$_2$Ti$_{1-x}$Mn$_x$Sn compositions. The local lattice distortion in the prepared compositions leads to weak random anisotropy, affecting the overall magnetic properties.

\section{Acknowledgment}
 K.M. acknowledges DST-INSPIRE (File: DST/INSPIRE/03/2021/001384/IF190777), New Delhi, for providing the research fellowship. KM and PB thank Dr. Mukul Gupta, UGC-DAE CSR, Indore, for extending the  X-ray diffraction facility. Acknowledgment is also due to Prof. J. G. Lin, Center for Condensed Matter Science, National Taiwan University, Taipei, Taiwan, for providing access to the magnetic measurement facility. We also thank Prof. Pratap Raychaudhuri and Ganesh Jangam for their help in magnetic measurement at the Tata Institute of Fundamental Research, Mumbai. K.M. thanks Mr. Rajeev Joshi, UGC-DAE CSR, for helping with electrical transport measurements. Part of this research is carried out at the light source PETRA III at DESY, a member of the Helmholtz Association (HGF). KM and PB thank Dr. Edmund Welter for his assistance in using the photon beamline P65. We acknowledge the financial support from the Department of Science and Technology (Govt. of India) provided within the framework of the India@DESY collaboration).

\bibliography{mybib}
\end{document}